\documentclass[ reprint, amsmath,amssymb, aip]{revtex4-2}
\usepackage{graphicx}
\usepackage{color}
\usepackage[colorlinks=true, allcolors=blue]{hyperref}
\usepackage{braket}
\usepackage{overpic}

\begin{document}

% \title{Detection of HF and VHF Fields through Floquet Sideband Gaps in Dressed Rydberg Atoms}
\title{Detection of HF and VHF Fields through Floquet Sideband Gaps by `Rabi Matching' Dressed Rydberg Atoms}
% Pseudo-resonant Detection of HF and VHF Fields by Rydberg Atoms
\author{Andrew P. Rotunno}
\author{Samuel Berweger}
\author{Nikunjkumar Prajapati}
\author{Matthew T. Simons}
\author{Alexandra B. Artusio-Glimpse}
\author{Christopher L. Holloway}
\thanks{christopher.holloway@nist.gov}
\affiliation{National Institute of Standards and Technology, Boulder, CO 80305, USA}
\author{Maitreyi Jayaseelan}
\affiliation{Department of Physics, University of Colorado, Boulder, Colorado 80302, USA}
\author{R. M. Potvliege}
\author{C. S.  Adams}
\affiliation{Department of Physics, Durham University, South Road, Durham DH1 3LE, United Kingdom}
\date{\today}

\begin{abstract}
Radio frequencies in the HF and VHF ($3$~MHz to $300$~MHz) bands are challenging for Rydberg atom-based detection schemes, as resonant detection requires exciting the atoms to extremely high energy states. We demonstrate a method for detecting and measuring radio frequency (RF) carriers in the HF and VHF bands via a controlled Autler-Townes line splitting. Using a resonant, high-frequency (GHz) RF field, the absorption signal from Townes-Merrit sidebands created by a low frequency, non-resonant RF field can be enhanced. Notably, this technique uses a measurement of the optical frequency separation of an avoided crossing to determine the amplitude of a non-resonant, low frequency RF field.
This technique also provides frequency-selective measurements of low frequency RF electric fields. To show this, we demonstrate amplitude modulated signal transduction on a low frequency VHF carrier. We further demonstrate reception of multiple tones simultaneously, creating a Rydberg `spectrum analyzer' over the VHF range. 
\end{abstract}
\maketitle

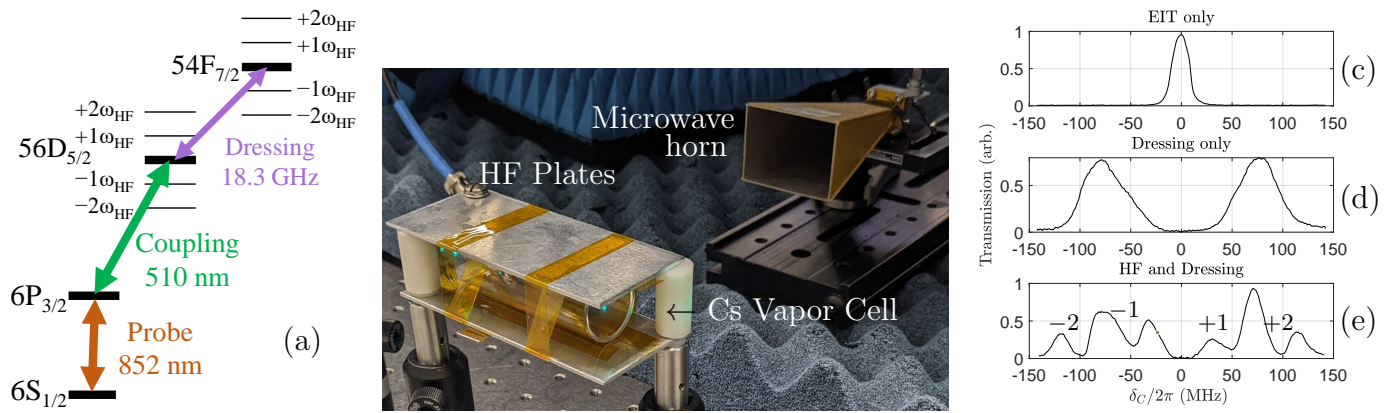
\begin{figure*}
    \centering
    \begin{overpic}[width=.27\textwidth]{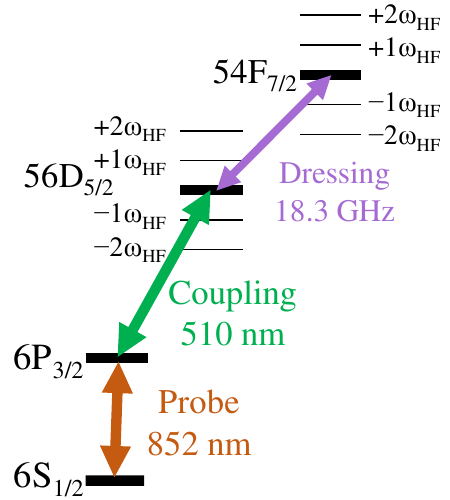}
    \put(70,15){\large (a)}\end{overpic}\hfill
    \begin{overpic}[width=.42\textwidth]{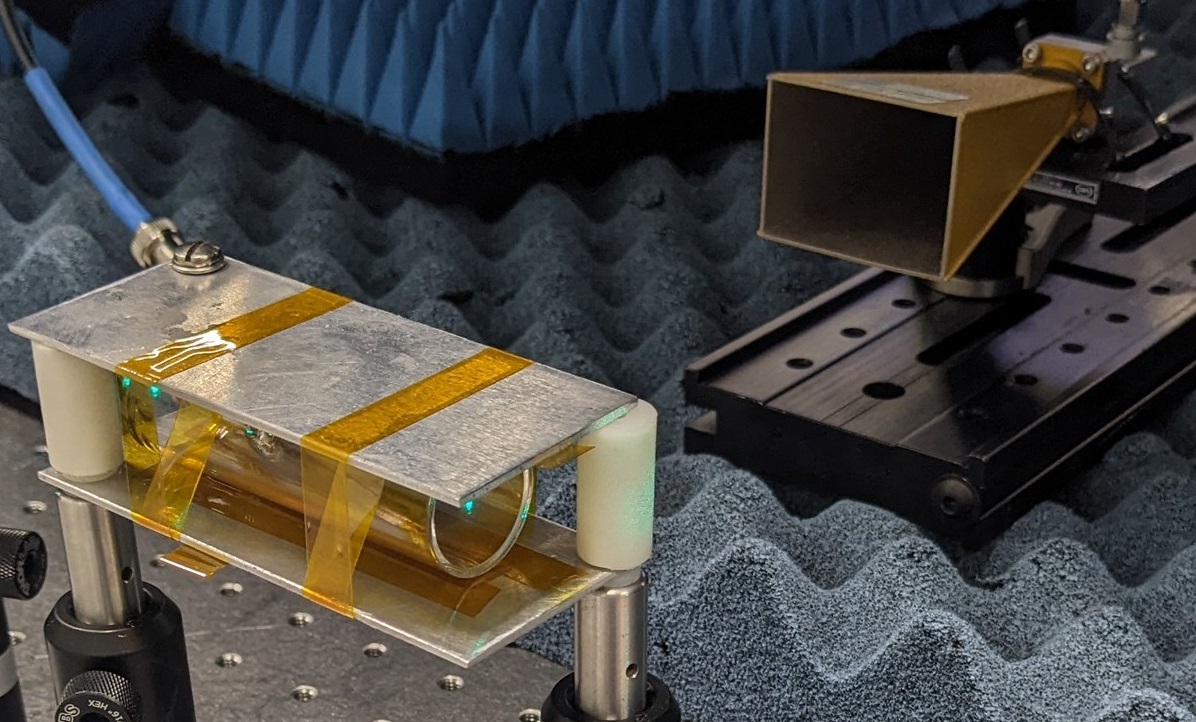}
    \put(17,40){\large \color{white} HF Plates}
    \put(37,50){\large \color{white} Microwave}
    \put(50,45){\large \color{white} horn}
    \put(50,16){\large \textcolor{black}{$\leftarrow$} \color{white}  Cs Vapor Cell}
    \end{overpic}\hfill
    \begin{overpic}[width=0.28\textwidth]{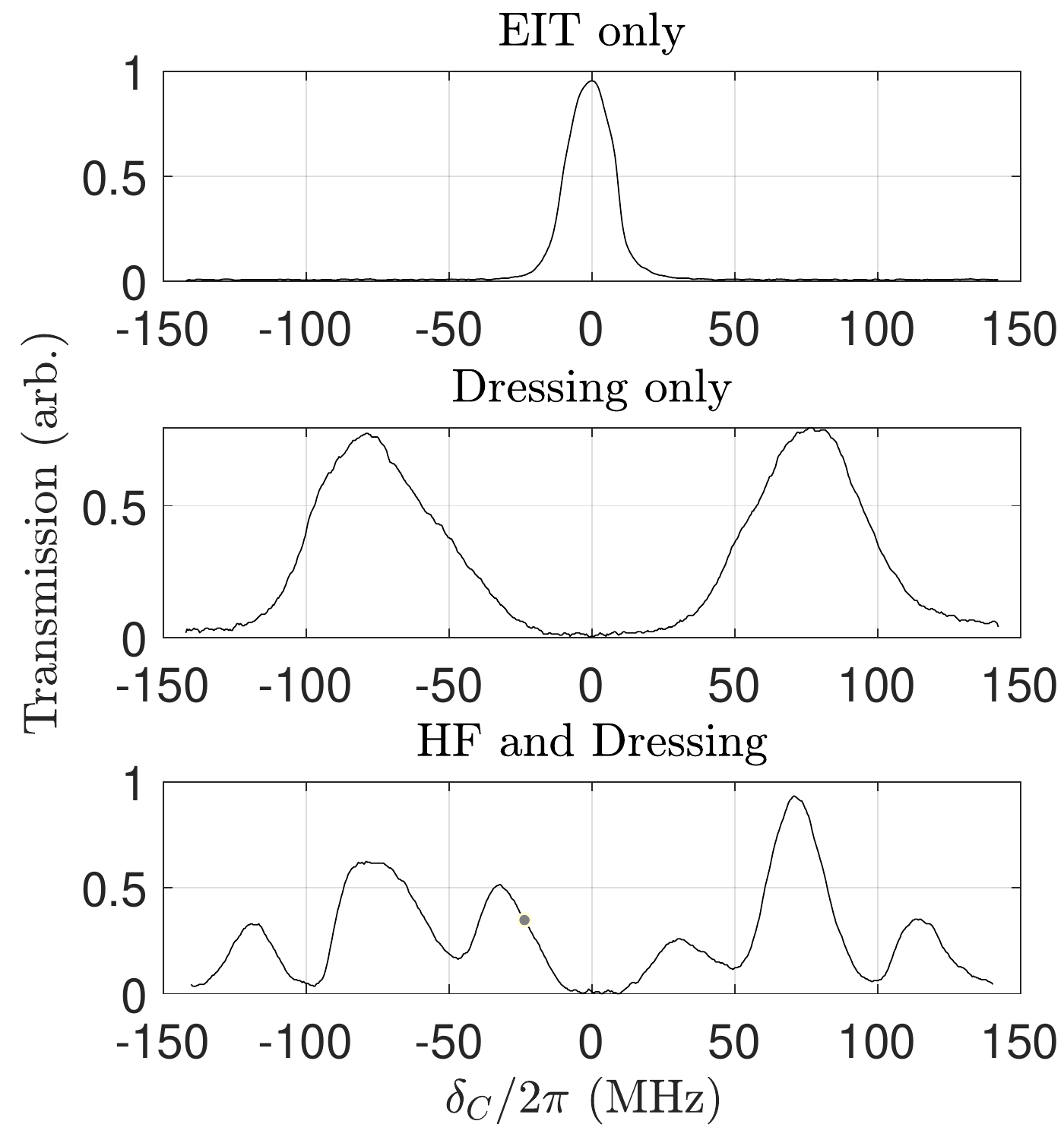}
    \put(91,82){\large (c)} 
    \put(91,50){\large (d)} 
    \put(91,20){\large (e)} 
    \put(17,20){ $-2$} 
    \put(32,23){ $-1$} 
    \put(54,20){ $+1$} 
    \put(70,20){ $+2$} 
    \end{overpic}
    \caption{     \label{fig:money}
    Overview of the experiment.
    (a) Atomic level diagram, 
    (b) Experimental setup, 
    and sample transmittance curves comparing (c) unperturbed EIT, (d) the dressed AT splitting with $\Omega_{dr}\approx2\pi\times150$~MHz, and (e) with the HF field ($E_{HF}\approx0.14$~V/cm) at 100~MHz (blue).
    The two pairs of observed line splittings are labelled $\pm1$ and $\pm2$. The Stark offset has been removed for illustration.}
\end{figure*}

\section{Introduction}
Highly excited Rydberg atomic states \cite{gallagher2006rydberg, adams2019rydberg} have been used as self-calibrated sensors for both dc \cite{osterwalder1999using,mohapatra2007coherent, holloway2022electromagnetically} and resonant radio frequency (RF) \cite{sedlacek2012microwave, holloway2014sub, artusio2022modern, meyer2020assessment} electric fields, due to the large dc polarizability of the higher orbital angular momentum states \cite{zimmerman1979stark, zhang1994measurement} and the large transition dipole moments of transitions between nearby Rydberg states. 
Atomic energy levels are typically probed spectroscopically via electromagnetically-induced transparency (EIT).
RF and dc electric fields are measured by changes in the EIT spectrum, with values and uncertainties based on physical atomic quantities.  
Measurements of laser transmittance alone are subject to many experimental sources of noise, but frequency-space measurements of atomic states using a scanning laser provide a more robust measurement.

Autler-Townes (AT) splitting caused by resonant RF fields provides measurement of GHz to THz fields, traceable to the International System of Units (SI)\cite{holloway2014broadband}, often with excellent accuracy.
One advantage to using Rydberg atoms for RF field sensing is their broad frequency range of operation. This is enabled by the spectral density of Rydberg states, with many large electric dipole moment transitions with resonant frequencies from MHz to THz.
However, Rydberg transitions in the HF and VHF range of 3~MHz~to~300~MHz require either very high principal quantum number states ($n > 100$), or high orbital angular momentum states \cite{ thaicharoen2019electromagnetically, brown2022vhf}. Either method suffers from weaker optical couplings for EIT, as well as contamination of the signal from the very large density of nearby atomic states.
Off-resonant generalized Rabi measurements are possible using nearby existing resonances \cite{simons2016using}, or farther away using ac Stark shifts with a strong local oscillator (LO) \cite{hu2022continuously}. A dc field can also be used to shift the resonance, though this may require a very strong and spatially homogeneous dc field.
%This work targets a range in the HF and VHF radio bands (we will henceforth use HF and VHF interchangeably for the purpose of this paper), which is `low' relative to the typically GHz-range Rydberg transitions. 
%Resonant Rydberg transitions the HF and VHF range of 3~MHz~to~300~MHz require either high principle quantum number which hinders intermediate transition coupling, or additional photons to reach higher orbital angular momentum states \cite{ thaicharoen2019electromagnetically, brown2022vhf}.
Power and frequency tuning methods  \cite{simons2021continuous, berweger2022rydberg, bohlouli2007enhancement} exist for shifting resonances to meet an arbitrary field, but these require signals near existing resonances.
The ability to detect low frequency or long-wavelength signals would enable long-distance reception in a compact form factor, with the active receiver volume only a few cm$^3$, compared to meters or longer classical dipole antennas \cite{cox2018quantum, meyer2020assessment, bussey2022quantum}, enabling compact long-distance reception for over-the-horizon, underground, or underwater applications.

Here we present a method for the detection of arbitrary electric fields in the 3~MHz to 300~MHz range (we will henceforth use HF and VHF interchangeably for the purpose of this paper) using a Rydberg state with no nearby HF or VHF transitions. 
We use a high-frequency RF field that is resonant with a nearby Rydberg transition to `match' the Rabi frequency with an applied HF signal field. Tuning the Rabi frequency of the `matching' field effectively probes the Floquet sideband structure generated by the HF field.
Absorption dips within the AT peaks reveal an avoided energy crossing, at the mid-point of the Floquet quasi-energy sidebands and the central EIT line.
The generation of the Floquet sidebands can be referred to as the Townes-Merritt (TM) effect \cite{townes1947stark, white1989theoretical, tanasittikosol2011rydberg, autler1955stark}, where cyclic modulation of state's Stark shift energy with a MHz-range field will produce quasi-energy sidebands on the observed EIT line  \cite{bason2010enhanced, miller2016radio, jiao2016spectroscopy, miller2017optical, jiao2017atom,  paradis2019atomic}.
This effect has been analyzed in other quantum systems as well as Rydberg EIT \cite{clark2019interacting, son2009floquet, hausinger2010dissipative}.

% When measuring these fields using highly polarizable Rydberg states, the applied root-mean-squared (RMS)  and background dc field both Stark shift the state's energy, moving the EIT line and the state's microwave resonances to nearby states. 

%This work demonstrates a method to transform measurements of electric field intensity from relative population measurements in TM sidebands into frequency-space splittings, measuring MHz-range electric field intensities spectrally, using an arbitrary Rydberg state. 
To probe the TM or `Floquet' sidebands caused by an HF field, we drive a simultaneous AT splitting, which splits both the center EIT peak and the sidebands.
When we `match' the dressing field's Rabi frequency to the applied HF, an avoided level crossing between the splitting EIT line and its sidebands is observed. The frequency gap grows non-linearly with the HF field strength.
%In this work, we present spectral measurements alongside a simplified two-level theory to obtain field amplitude measurements.
%Despite the current lack of precision, this demonstration of a spectral field measurement for dc and MHz-range electric fields opens a path toward precision measurement, given a many-sate atomic model and better microwave field uniformity than this proof-of-principle demonstration. 
We briefly illustrate the effect in Fig.~\ref{fig:money}, showing (a) an atomic energy level diagram, and (b) the experimental setup, and in 
(c-e) we show transmittance spectra demonstrating the effect of a 100~MHz HF field on microwave-driven EIT-AT spectrum.
We observe both first and second order sidebands in transmittance signal at $\pm$50~MHz and $\pm$100~MHz detuning from the central EIT line.
With appropriate atomic analysis, the non-linear HF field splitting gives measurements of electric field ac amplitude and dc offset \cite{y2006parity}, as well as the frequency via the spectrum, and phase can be measured with an LO.
The splitting allows for a measurement range continuous from a low end of the EIT Doppler linewidth to an upper roll-off with broadening caused by field non-uniformity \cite{rotunno2022modeling}, roughly 5~MHz to 500~MHz in our demonstration.
In principle, the range can be extended up to GHz with uniform electric fields, and sub-MHz with Doppler cancellation or cold atoms. 
%We find this method is less sensitive for communication than simple Stark-shifting schemes \cite{meyer2021waveguide, liu2022highly}, although there is an inherent optical linewidth band-pass, by using the Rabi frequency as a pseudo-LO in our scheme. %<---- put this in the analysis, not here.

We discuss our experimental setup in Sec.~\ref{sec:Exp}, and present a combination of Floquet and dressed atom theory in Sec.~\ref{sec:The} and Appx.~\ref{appx:theory}.
We show results of experimental data against theory and the effect of multiple simultaneous HF fields in Sec.~\ref{sec:Res}.
Looking toward applications, we demonstrate the ability to perform `pseudo-resonant' HF detection in Sec.~\ref{sec:signals}, and conclude in Sec.~\ref{sec:conc}.
In Appx.~\ref{appx:abbrev}, we list the non-standard symbols used in this paper.

\section{Experimental Setup}\label{sec:Exp}
The Rydberg EIT-AT setup is shown in the level diagram of  Fig.~\ref{fig:money}(a) and image of the experiment Fig.~\ref{fig:money}(b), including two counter-propagating lasers, a cesium vapor cell (75 mm $\times$ 25.4 mm $\oslash$ ) between two parallel plates, and a K-band microwave horn placed approximately 15~cm from the atoms. 
The 852~nm probe laser is held near resonance on the D2 transition $\bra{6S_{1/2}, F=4} \leftrightarrow\ket{6P_{3/2}, F=5}$. The probe transmittance is measured through differential detection, with and without the overlapping 510~nm coupling laser. The coupling laser frequency detuning $\delta_C$ from the $\ket{56D_{5/2}}$ state EIT resonance is scanned for an EIT spectrum measurement.
The probe beam had a ($1/e^2$) waist diameter of 1.03~mm and a power of $0.5$~mW, yielding a probe Rabi frequency of $\approx~2\pi\times~31.5$~MHz. The coupling beam had a waist diameter of 1.7~mm and a power of 33~mW, yielding a coupling Rabi frequency of $\approx~2\pi\times~0.9$~MHz \cite{vsibalic2017arc}. 
In this letter we plot the differential probe transmittance intensity as a false-color axis while coupling laser detuning $\delta_C/2\pi$ is scanned on the horizontal axis, averaged over 5 traces in the oscilloscope before read-out.
Scanning either HF or RF power along the vertical axis gives `waterfall' spectrum plots of experimental data that we compare against theory. 
Long-term drifting offsets of the coupling laser are corrected using a simultaneous field-free reference cell, and the frequency scale is given by the low-field limit of the fine energy gap between the $\ket{56D_{3/2}}$ and $\ket{56D_{5/2}}$ states (392~MHz \cite{vsibalic2017arc}).

The resonant GHz dressing field is applied using a horn antenna, also shown in Fig.~\ref{fig:money}(b).
Given the long wavelength of the HF and VHF bands, instead of using a large antenna for propagation, we apply these RF fields across two parallel plates (shown in Fig.~\ref{fig:money}(b))~\cite{jiao2017atom, miller2016radio, liu2022highly, paradis2019atomic, jiao2016spectroscopy, miller2017optical, jiao2017atom}. 
Each plate is 102~mm$\times$45~mm in size, supplied electrically by stranded wires screwed down at tapped holes near the corner, and separated by two 25.4~mm insulating dowels on opposite corners. 
%The plates are poorly impedance matched to the incoming power $P_{HF}$, and we record the empirical amplitude lost in the reflected $S_{11}$ parameter as the dashed line in Fig.~\ref{fig:signals}(d).
While the impedance of the glass cell is unknown at low frequency, in theory the glass vapor cell insulates against external dc-limit electric fields \cite{jau2020vapor}.
However, we observe a residual dc or root-mean-squared (RMS) electric field in the following atomic measurements, possibly due to internal residual or laser-induced charges \cite{bason2010enhanced}, or background radio waves.

The Rydberg transition driven by the dressing field is illustrated in Fig.~\ref{fig:money}(a), on $\braket{56D_{5/2}, m_J=\frac{1}{2}|\leftrightarrow|54F_{7/2}, m_J=\frac{1}{2}}$, which has a dipole transition strength of 1746~$ea_0$, and transition frequency measured near 18.313(1)~GHz, significantly different than the expected resonance at 18.346~GHz \cite{vsibalic2017arc}, indicating a potential stray field in the cell.
This applied frequency $\omega_{dr}$ must compensate for the differential Stark shifting of both states with the HF power, $P_{HF}$.
This effective resonance is determined by hand-tuning $\omega_{dr}$ at MHz intervals to obtain equal AT peaks in the low-field limit, then the high-field limit when the HF is applied. 
This shift was measured empirically at roughly $-579(14)$~MHz per mW applied at 100~MHz, and using the power-to-field conversion determined in Sec.~\ref{sec:Res}, nearly $-3.62(9)$~GHz per (V/cm)$^2$, about $4\%$ below the theory value from the difference in polarizabilities (see Appx.~\ref{appx:theory}).
This power conversion ($\sqrt{P_{HF}}$ to $|E_{HF}|$) is measured at 100~MHz, and in general depends on the frequency-dependent coupling of the plates and vapor cell. 
In order to make this conversion, we require an atomic theory for the measurement of $E_{HF}$ using the observed splittings.

% $-578.9~\pm6.9$~MHz per mW !convert * 0.42^2 V/cm per 1mW!

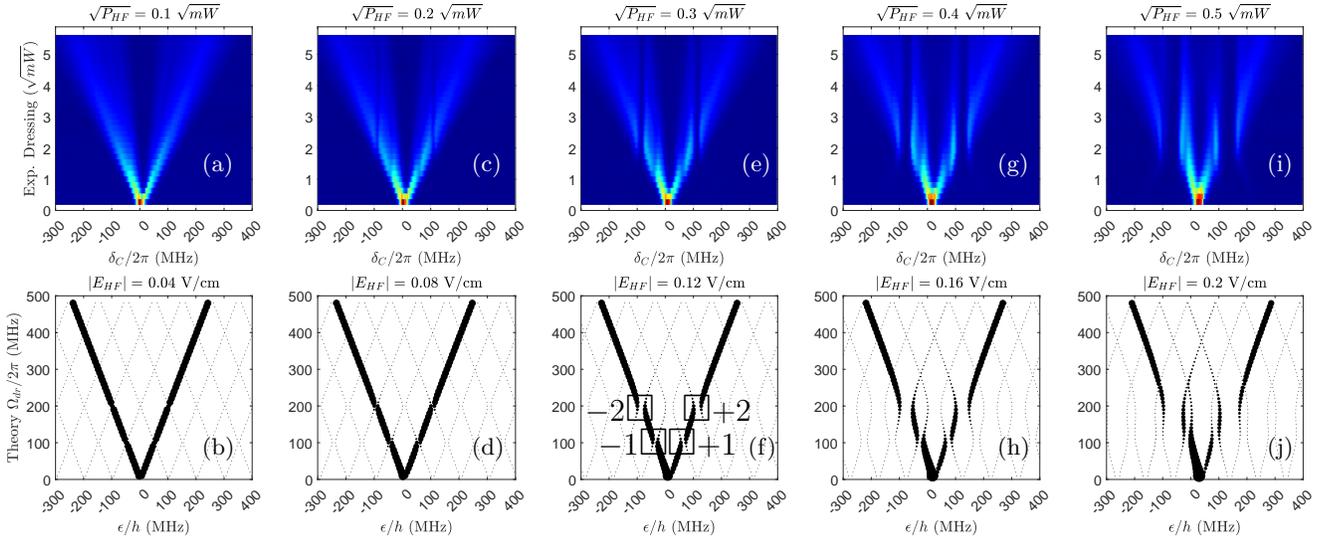
\begin{figure*}
    \centering
    \begin{overpic}[width=\textwidth]{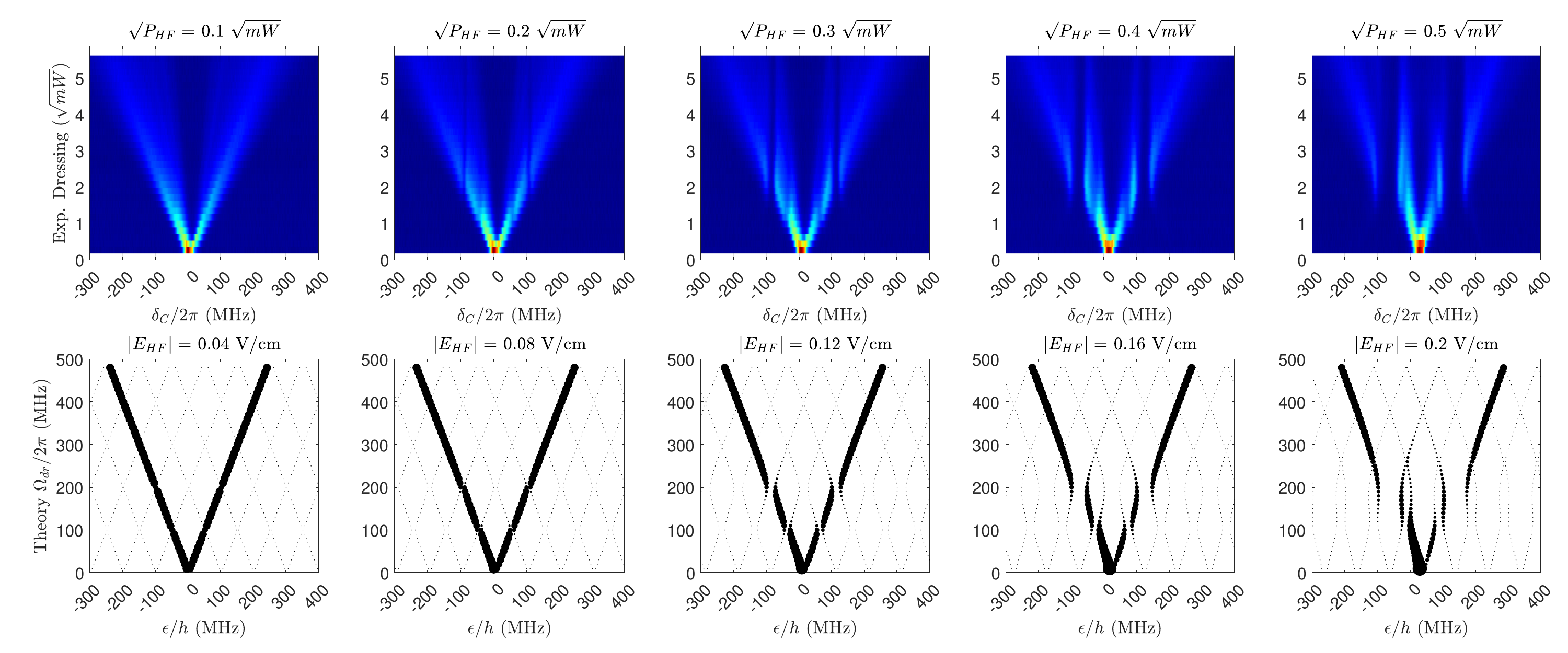}
    \put(16,28){ \textcolor{white}{(a)}   }
    \put(36,28){ \textcolor{white}{(c)}   }
    \put(56,28){ \textcolor{white}{(e)}   }
    \put(75,28){ \textcolor{white}{(g)}   }
    \put(95,28){ \textcolor{white}{(i)}   }
    \put(16,7){ \textcolor{black}{(b)}   }
    \put(36,7){ \textcolor{black}{(d)}   }
    \put(56.5,7){ \textcolor{black}{(f)}   }
    \put(75,7){ \textcolor{black}{(h)}   }
    \put(95,7){ \textcolor{black}{(j)}   }
    \put(44.4,9.5){ {\large ${-2}$}{\Large $\square$}} %\large $^\sim$
    \put(46.0,7){{\large ${-1}$}{\Large $\square$} } %\large $\times$
    \put(51.1,7){{\Large $\square$}{\large ${+1}$}} %\large $\times$
    \put(52.2,9.5){{\Large $\square$}{\large ${+2}$}}\end{overpic} %\large $^\sim$
    \caption{Comparison of experiment and theory, with the two pairs of avoided crossings labelled $\pm1, \pm2$. 
    Top: Experimentally observed transmittance spectra showing a waterfall over applied dressing field vertically, with each plot from left to right increasing amplitude ($\sqrt{P_{HF}}$) of a 100 MHz HF field. 
    Bottom: Calculated state quasi-energy $\varepsilon$, with state projection of $\ket{56D_{5/2}}$ as marker size, plotted as a corresponding waterfall over $\Omega_{dr}$.
    Theory plots use $E_{dc}=0.02$~V/cm, $N_{max}=24$, $S_\delta = 1.1$, scaling $\Omega_{dr}/\sqrt{P_{dr}} = 85\times2\pi$~MHz~$/\sqrt{\textrm{mW}}$, varying $\Omega_{dr}$ and $E_{HF}$ as labeled.
    \label{fig:ScanDressing}}
\end{figure*}

\section{Theory}\label{sec:The}
%We note this project began as an attempt to `split the splitting,' or drive a typical two-level dressed atom transition \textit{between} dressed AT states, when the splitting or Rabi frequency $\Omega_{dr}$ is nearly matched by a second field $\omega_{HF}$. 
We employ a model where the Rabi frequency $\Omega_{dr}$ of the dressing field drives resonant transitions between two Floquet `ladders' of quasi-energy states, as illustrated in Fig.~\ref{fig:money}(a).
%We find that the observed effect cannot be interpreted as a simple linear `second' splitting of the dressed Rydberg states, due to the apparent non-linear splitting by the HF field, and the appearance of two pairs of HF resonances.
We differentiate the method used here, namely the Floquet expansion of a dressed two-level system,  from other methods including the use of a modulated RF field \cite{meyer2018digital, anderson2020atomic, holloway2020multiple, zou2020atomic, liu2022deep, borowka2022sensitivity}, multiple tones near Rydberg resonances \cite{ kim2000semiclassical, anderson2014two, anderson2016optical, anderson2017continuous, xue2021microwave, ficek1999quantum, yu1997driving}, leverage off-resonant ac Stark tuning using a resonance \cite{zou2020atomic, yang2022wideband}, use continuous resonance tuning methods \cite{simons2021continuous, berweger2022rydberg}, or observe Floquet (TM) sidebands alone \cite{townes1947stark, bason2010enhanced, miller2016radio, jiao2016spectroscopy, chai2021demonstration}. 
Analysis for two-level physics where the states are dressed and driven have been performed \cite{ tanasittikosol2012sidebands, son2009floquet, hausinger2010dissipative, shirley1965solution}, and the effect can be extended to other driven two-level quantum systems. 

We focus on the dynamics of a two-state Rydberg system, leaving aside details of the two-photon cascade EIT which populates Rydberg states and probes energy levels spectrally \cite{mohapatra2007coherent}.
We consider the states $\ket{56D_{5/2},m_J=\frac{1}{2}}=\ket{D}$ and $\ket{54F_{7/2},m_J=\frac{1}{2}}=\ket{F}$ in Cesium, which are subject to three relevant electric fields at different rates, simultaneously, applied perpendicular to the table:
1) the dressing field is nearly resonant to the strong D-F Rydberg transition $\omega_{dr}\simeq \omega_F-\omega_D$, performing controlled AT splitting, 2) an HF field of interest at a much lower frequency $\omega_{HF} \ll \omega_F-\omega_D$ which populates Floquet sidebands in the EIT spectrum, and 3) a dc or off-resonant effective RMS background field which exists in the cell.

In the low-field limit, the Stark energy shift due to the HF field evolves over time as $-\alpha_{D/F}\mathcal{E}(t)^2/2$, proportional to electric field squared $\mathcal{E}(t)^2$, and dc polarizabilities $\alpha_{D/F}$, where the subscript  $(~\cdot~)_{D/F}$ represents throughout the value for either individual state.
Proper Stark shift calculations are required at higher fields, considering contributions from many allowed dipole transitions which couple to the electric field, contributing to total energy shift and population mixing.
We show polarizability calculations for $\ket{D}$ and $\ket{F}$ in Appx.~\ref{sec:pol}.
We fit frequency shift over $\mathcal{E}^2$ for effective polarizability in the field range used in this work.
Throughout this analysis we use $\alpha_D=-3003$~MHz per (V/cm)$^2$ and $\alpha_F=12100$~MHz per (V/cm)$^2$.
We utilize an effective polarizability for its computational simplicity, although one could use a `multi-state' computation using relevant states from the Stark calculation for more accuracy, especially for high fields. 
For instance, the $\ket{F}$ state represents $<90\%$ of the population for only $0.008$~(V/cm)$^2$, shown in Appx.~\ref{sec:pol}. 
We see good agreement between a multi-state model and the one employed here over the modest field ranges used, and we plan to investigate features which are only captured by a multi-state model in future work.
% re: robert's comment about talking about the multistate

The electric field of our HF sinusoid and a dc offset can be written $ \mathcal{E}(t) = E_{dc} + E_\textrm{HF}\cos(\omega_{HF}t) $.
The $E_{dc}$ term can be considered to contain broadband background ac electric fields, effectively an RMS contribution to the average dc field value.
Since the Stark shift $-\alpha_{D/F}|\mathcal{E}(t)|^2/2\hbar$ is proportional to field squared, we use the following abbreviations for the ac ($\sim$), dc ($-$), and cross-term ($\times$) components:
\begin{equation}\label{eq:sigmadef}
    \begin{split}
        \Sigma_{D/F}^\sim=&\frac{-\alpha_{D/F}E_{HF}^2}{4\hbar} \\
        \Sigma_{D/F}^-=&\frac{-\alpha_{D/F}E_{dc}^2}{2\hbar} \\
        \Sigma_{D/F}^\times=&\frac{-\alpha_{D/F}E_{dc}E_{HF}}{\hbar} 
    \end{split}
\end{equation}
letting us write the time-evolving Stark shifts:
\begin{equation}\label{eq:terms}
\begin{split}
    \frac{-\alpha_{D/F}|\mathcal{E}(t)|^2}{2\hbar} = \hspace{5pt} &\Sigma_{D/F}^- + \Sigma_{D/F}^\sim \\
    +&\Sigma_{D/F}^\times \left( e^{i\omega_{HF}t} + e^{-i\omega_{HF}t} \right)/2 \\ 
    +&\Sigma_{D/F}^\sim  \left( e^{i2\omega_{HF}t} + e^{-i2\omega_{HF}t} \right)/2 
\end{split}
\end{equation}

When time-averaging, only the first two terms remain, the dc and HF RMS fields, which define our time-averaged shifted resonance from $\omega_{dr}\simeq\omega_F-\omega_D$. 
That is, due to a differential Stark shift from the HF field intensity, an applied dressing field must shift in order to observe equal AT splitting of the EIT line, with a new effective detuning between the applied field $\omega_{dr}$ and the shifted resonance: $\delta_{dr}\equiv \omega_{dr}-\left[\left(\omega_{F} + \Sigma_{F}^- + \Sigma_{F}^\sim\right) - \left(\omega_{D} + \Sigma_{D}^- + \Sigma_{D}^\sim\right)\right]$.
We adjust the applied value of $\omega_{dr}$ to keep $\delta_{dr}\simeq0$ in the high-field limit, by optimizing the symmetry of spectral peaks. 
We also invoke a detuning scaling parameter to compensate for the difference in experimental resonance shift and the theory values from the effective polarizabilities: $\omega_{dr}-(\omega_F-\omega_D)=S_\delta\cdot(\Sigma_{F}^- + \Sigma_{F}^\sim-\Sigma_{D}^- - \Sigma_{D}^\sim)$, where the scaling parameter $S_\delta=1$ gives the theoretical resonance ($\delta_{dr}=0$).

Aside from constant terms, there are oscillating components: a dc-inclusive cross term at $\pm1\omega_{HF}$ with strength $\Sigma_{D/F}^\times/2\propto E_{dc}E_{HF}$, and an ac `intensity' term at $\pm2\omega_{HF}$ with strength $\Sigma_{D/F}^\sim/2\propto E^2_{HF}$.
The rectified oscillating field $\cos^2(\phi)=\frac{1}{2}\left(1+\cos(2\phi)\right)$  doubles the effective modulation frequency, making even quasi-energy sidebands with multiples of $2\omega_{HF}$ from the center EIT line \cite{paradis2019atomic}. 
The dc field breaks this symmetry, enabling single $\omega_{HF}$ photon transitions, thus odd sidebands in multiples of $\omega_{HF}$ \cite{miller2016radio}.
For the case of two applied HF frequencies, we expect ac components which oscillate at sum and difference frequencies as  $\cos(\phi)\cos(\theta)=\frac{1}{2}\left[\cos(\phi-\theta)+\cos(\phi+\theta)\right]$, which are demonstrated in Sec.~\ref{sec:Res}.

% The first two terms are time-independent Stark shifts, proportional to $|E_{HF}|^2/2$ and $E_{dc}^2$, then rotating components which oscillate at $\omega_{HF}$ with strength $\Sigma_{D/F}^\times/2$  (the dc-inclusive cross term), and at $2\omega_{HF}$ with strength $\Sigma_{D/F}^\sim/2$ (the pure ac term). 

We discuss the creation of a time-independent Hamiltonian using both Floquet states and the dressed-atom approach in Appendix~\ref{appx:theory}.
Diagonalizing this Hamiltonian reveals an overall energy structure where states populate a `split ladder' of states with eigen-energies $\varepsilon_{N,\pm}\approx N\hbar\omega_{HF}\pm\hbar\Omega_{dr}/2$ around the original state energy $\varepsilon_0\approx\Sigma_D^-+\Sigma_D^\sim$, for integer $N$.
That is, on either side of the EIT resonance $\varepsilon_0$, we observe a first order pair of sidebands $\varepsilon_{\pm1}$ at $\varepsilon_0\pm\omega_{HF}$, a second order pair $\varepsilon_{\pm2}$ at $\varepsilon_0\pm2\omega_{HF}$, and so on.
Each quasi-energy ladder state splits into the ($\pm$) AT dressed states $\varepsilon_{N,\pm}$, approximately separated by Rabi frequency $\Omega_{dr}=\wp_{DF}E_{dr}/\hbar$ with dipole moment $\wp_{DF}=\bra{D}e\cdot\hat{z}\ket{F}$, as shown in Fig.~\ref{fig:ScanDressing}(a,b).
However, when these splitting sidebands approach each other (moving up the waterfall plot from the bottom), we observe an avoided energy level crossing which alters the state energies, and depends on both the HF and dressing fields, as seen in Fig.~\ref{fig:ScanDressing}(c-j) over increasing $E_{HF}$. 
We find that the observed effect cannot be interpreted as a simple linear second splitting of the dressed Rydberg states, due to the apparent non-linear splitting by the HF field, and the appearance of two pairs of HF resonances.

The characteristic feature of this method is observed as two pairs of avoided energy crossings in the spectrum for $\Omega_{dr}\simeq\omega_{HF}$ and $\Omega_{dr}\simeq2\omega_{HF}$ in the presence of an HF field $\omega_{HF}$.
The case of $\Omega_{dr}\simeq\omega_{HF}$ we refer to as the $\pm1$ feature (or $\varepsilon_{0,\pm} \simeq \varepsilon_{\pm1,\mp}$), and the case of $\Omega_{dr}\simeq2\omega_{HF}$ we call the $\pm2$ feature (or $\varepsilon_{0,\pm} \simeq \varepsilon_{\pm2,\mp}$). These splittings grow non-linearly with $\Sigma^\times_{D/F}/\omega_{HF}$ and $\Sigma^\sim_{D/F}/\omega_{HF}$, respectively. 
The spectral location of the avoided level crossing in the resonant case $\delta_{dr}=0$ (leaving aside the shift of $\Sigma_D^- + \Sigma_D^\sim$) is at $\pm\omega_{HF}/2$ for the first crossing and $\pm\omega_{HF}$ for the second crossing. These are at the midpoints between $\varepsilon_0$ and $\varepsilon_{\pm1}$, and $\varepsilon_0$ and $\varepsilon_{\pm2}$, respectively. 

To explain these additional avoided level crossings, we find that we can identify allowed transitions by considering the parity of the states involved and counting photons.
The $\pm1\omega_{HF}$ modulation drives an allowed transition between adjacent Floquet states of opposite (odd) parity, similarly for the $\pm2\omega_{HF}$ modulation allows a transition to the same (even) parity sideband. 
We note only the first and second order sideband crossings with the center $\varepsilon_0$ state are strongly avoided, while higher order crossings ($>2$) do not have a significant gap.

Deriving a closed-form expression for field values from a splitting measurement remains non-trivial, (i.e. not linear with field as in AT splitting), as the combination of both  dressing and HF fields shift the avoid energy crossings, and effective state polarizability depends on the range of $E_{HF}$.
This leaves us to diagonalize the Hamiltonian in Eq.~\ref{eq:bigHam} to find eigen-energies for various field values, and compare those to experimental field waterfall spectra, as in Sec.~\ref{sec:Res}. 
We did not attempt to calculate theory transmittance spectrum curves, although proper fitting would require this, and such theory could incorporate the range of Rabi frequencies sampled and Doppler effects, but is computationally expensive.

\begin{figure*}
    \centering
    \begin{overpic}[width=0.245\textwidth]{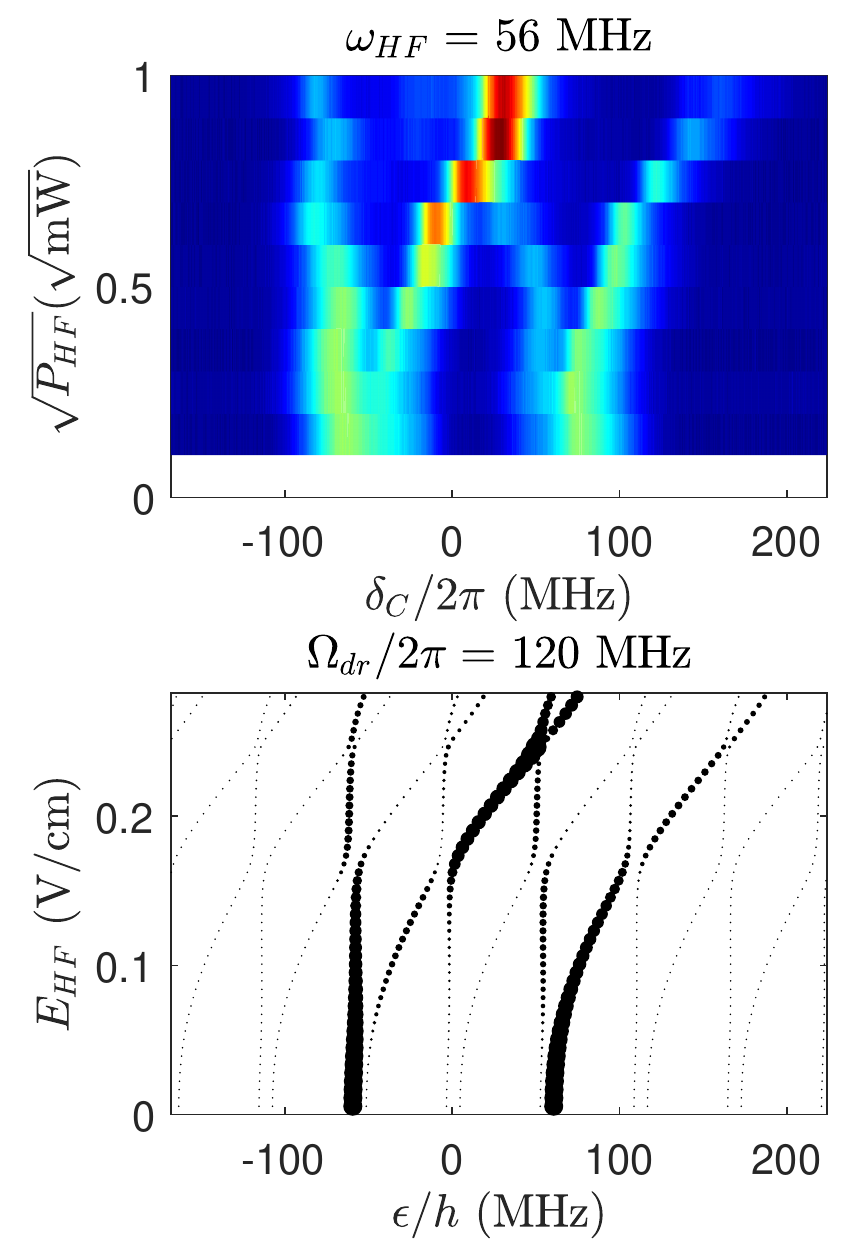}
    \put(1,93){(a)} \put(1,46){(b)} \end{overpic}
    \begin{overpic}[width=0.245\textwidth]{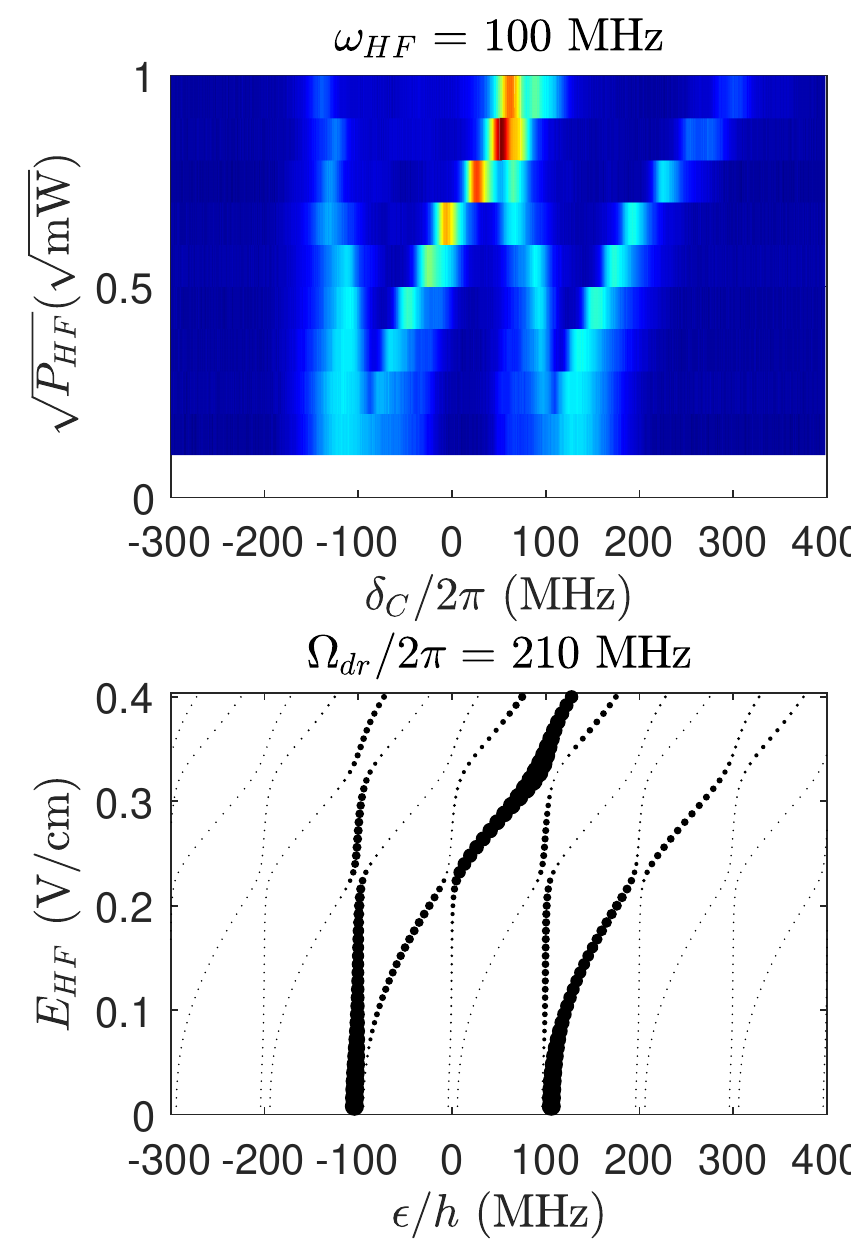}
    \put(1,93){(c)} \put(1,46){(d)} \end{overpic}
    \begin{overpic}[width=0.245\textwidth]{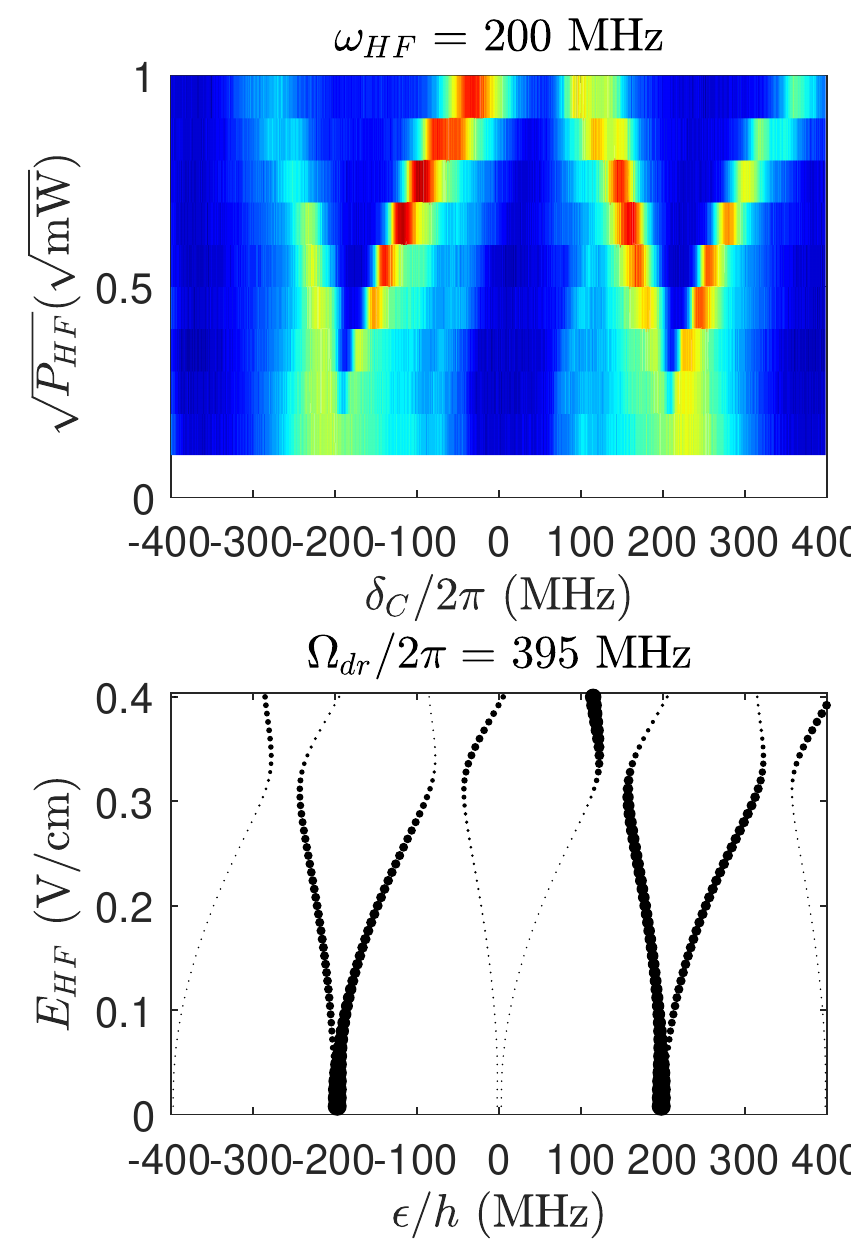}
    \put(1,93){(e)} \put(1,46){(f)} \end{overpic}
    \begin{overpic}[width=0.245\textwidth]{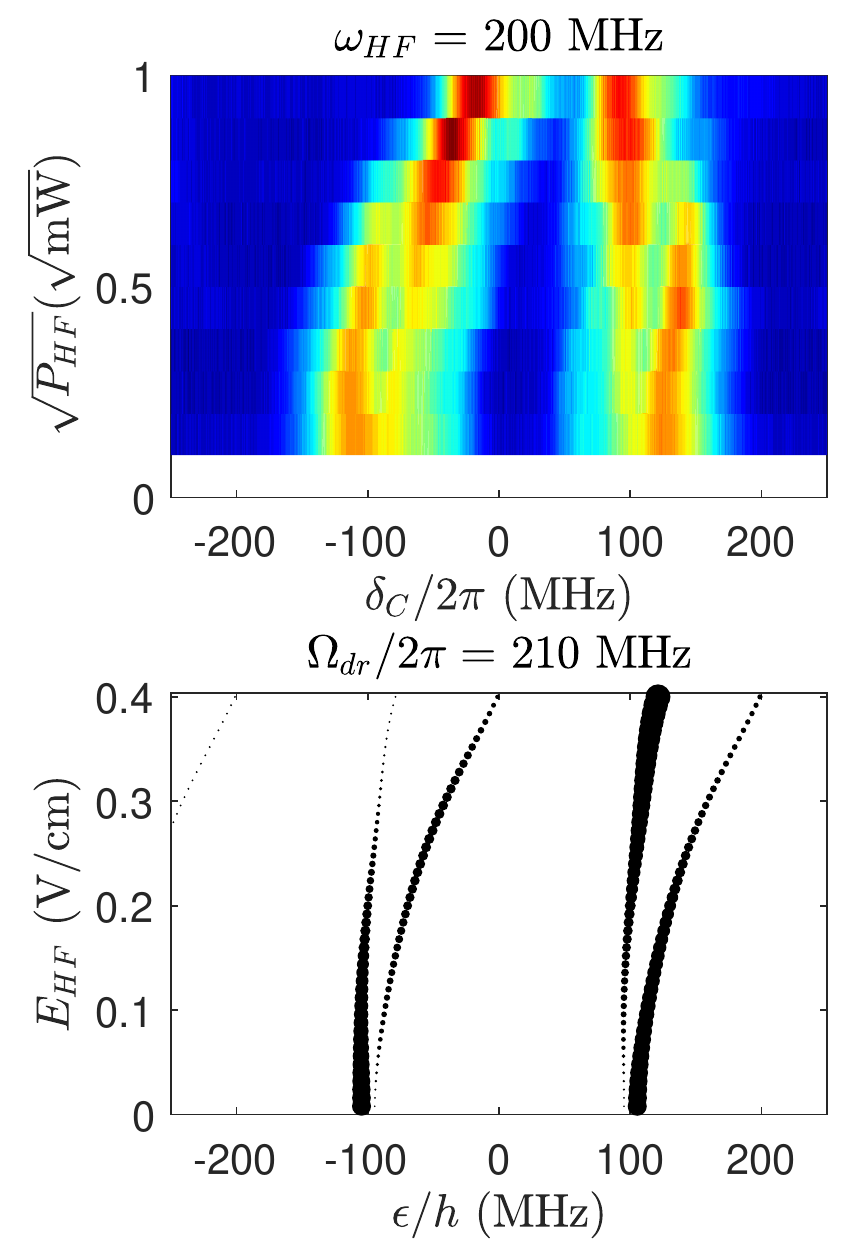}
    \put(1,93){(g)} \put(1,46){(h)} \end{overpic}
   \caption{Effect of increasing HF field, comparing experimental transmittance data (top) with theory quasi-energy (bottom). 
    We show $\omega_{HF}=$ (a,b) 56~MHz, (c,d) 100~MHz, (e-h) 200~MHz, using $\Omega_{dr}$ which demonstrate $\pm2$ (a-f), and $\pm1$ (g,h) splittings. 
    These plots hold $\Omega_{dr}$ constant, keeping experimental $\delta_{dr}\simeq0$ as $E_{HF}$ grows ($E_{dc}=20$~mV/cm), using  detuning scaling $S_\delta=$  (b) 1.05, (d) 1.1, (f) 0.8, and (h) 0.8.
    \label{fig:scanlowfreq}    }
\end{figure*}
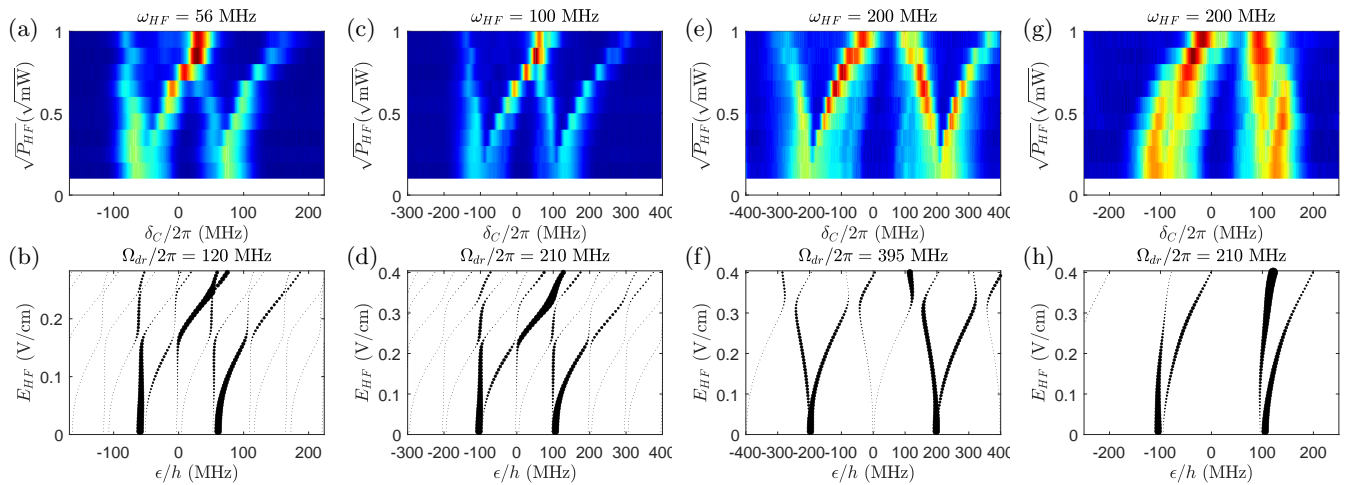

\section{Results}\label{sec:Res}
The primary finding here is the ability to detect far off-resonant HF and VHF signals using an induced line splitting, in order to convert a relative transmittance measurement to a more accurate frequency-space measurement. 
With control over laser detuning $\delta_C$ and dressing strength $\Omega_{dr}$, we can measure an arbitrary HF field's frequency, dc and ac components in a single spectrum. 
This provides a path to atomic measurements of field amplitudes over decades of frequency range (3~MHz to 300~MHz here), with around one decade of dynamic range demonstrated here.
A sample measurement is shown in Fig.~\ref{fig:money}, where transmittance spectra versus coupling laser detuning show (c) Rydberg EIT, (d) resonant dressing AT splitting, and (e) two pairs of absorption features, labelled $\pm1$ and $\pm2$.
Spectra like Fig.~\ref{fig:money}(e) for different RF powers are combined to create waterfall plots in following figures. 

In Fig.~\ref{fig:ScanDressing}, we compare experimental waterfalls over AT splitting  $\Omega_{dr}$ via controlled microwave power (top row), with calculated state quasi-energies (bottom row), illustrating the `$\pm1$' ($\Omega_{dr}\approx\omega_{HF}$) and `$\pm2$' ($\Omega_{dr}\approx2\omega_{HF}$) pairs of splittings. 
We vary $P_{HF}$ so that $E_{HF}$ changes linearly between columns, then adjust $\omega_{dr}$ so that $\delta_{dr}\simeq0$ in the high $\Omega_{dr}$ limit.
We use the overlay theory points on transmittance data to `fit-by-eye' to determine an input power to applied field conversion for the HF, the background field $E_{dc}$, and the dressing power-to-Rabi frequency conversion. 
Comparing theory to experiment yields an input power to applied field calibration factor $C_{PF}$  ($\sqrt{P_{HF}} = C_{PF} |E_{HF}|$) conversion at 100~MHz of $C_{PF} \approx 0.40$~V/cm per $\sqrt{\textrm{mW}}$. 
Due to the frequency-dependent coupling of the plates and the glass cell, we expect this conversion to change across applied $\omega_{HF}$.
We estimate a conservative error of order 10\%, given a combination of non-linear polarizability effects, EIT linewidth, and non-uniformity of $\Omega_{dr}$. % explain where the error is actually coming from - why 10%?
Additionally, the existence of a $\pm1$ sideband suggests residual dc field in the cell, estimated at around $E_{dc}=20$~mV/cm, using the size of the $\pm1$ splittings as $E_{HF}$ varies.
We also use this method to obtain the Rabi frequency conversion of $\Omega_{dr}\approx2\pi\times85$~MHz per $\sqrt{\textrm{mW}}$ of dressing power $P_{dr}$, using the low-field limit of HF in Fig.~\ref{fig:ScanDressing}(a,b).
% Is the Rabi in MHz or Rad/s?

In Fig.~\ref{fig:scanlowfreq}, we show spectra in waterfalls over $E_{HF}$ with fixed $\Omega_{dr}$, showing the non-linear induced splitting within each AT peak, as well as the Stark shifting of the $\ket{D}$ state with $E_{HF}^2$. 
Experimentally, $\Omega_{dr}$ is held constant ($\approx\omega_{HF}$ or $2\omega_{HF}$), but $\omega_{dr}$ is varied to maintain $\delta_{dr}\simeq0$ between rows. 
We do not compensate for frequency-dependent horn gain as we maintain signal generator output power across frequencies.
The comparison of theory with experiment shows characteristic agreement, which we expect to break down over the higher field values employed, as the real state energy shifts differ from the quadratic polarizability model employed.
%We note that the separation of the $\ket{56D_{3/2}}$ state and its movement with $E_{HF}$ in experimental observations deviate from even a multi-state theory, although these calculations are not presented here. 
The broadened lines in experimental AT spectra represent a sample over many values of a spatially non-uniform $\Omega_{dr}$ strength, while theory assumes a single value of $\Omega_{dr}$. 
As well, a constant $\alpha_{D/F}$ is employed over all sets, but in general this depends on $E_{HF}$ amplitude. 
There is significant interplay in the level shifts due to both fields at high power, causing difficulty in accurate modeling with simple two-level approximations, or obtaining closed-form expressions for the gap. 
Nevertheless, this measurement demonstrates a path toward making a spectral measurement of $E_{HF}$ and $E_{dc}$, using a controlled microwave dressing field. 

\begin{figure}
    \centering
    \begin{overpic}[width=\columnwidth]{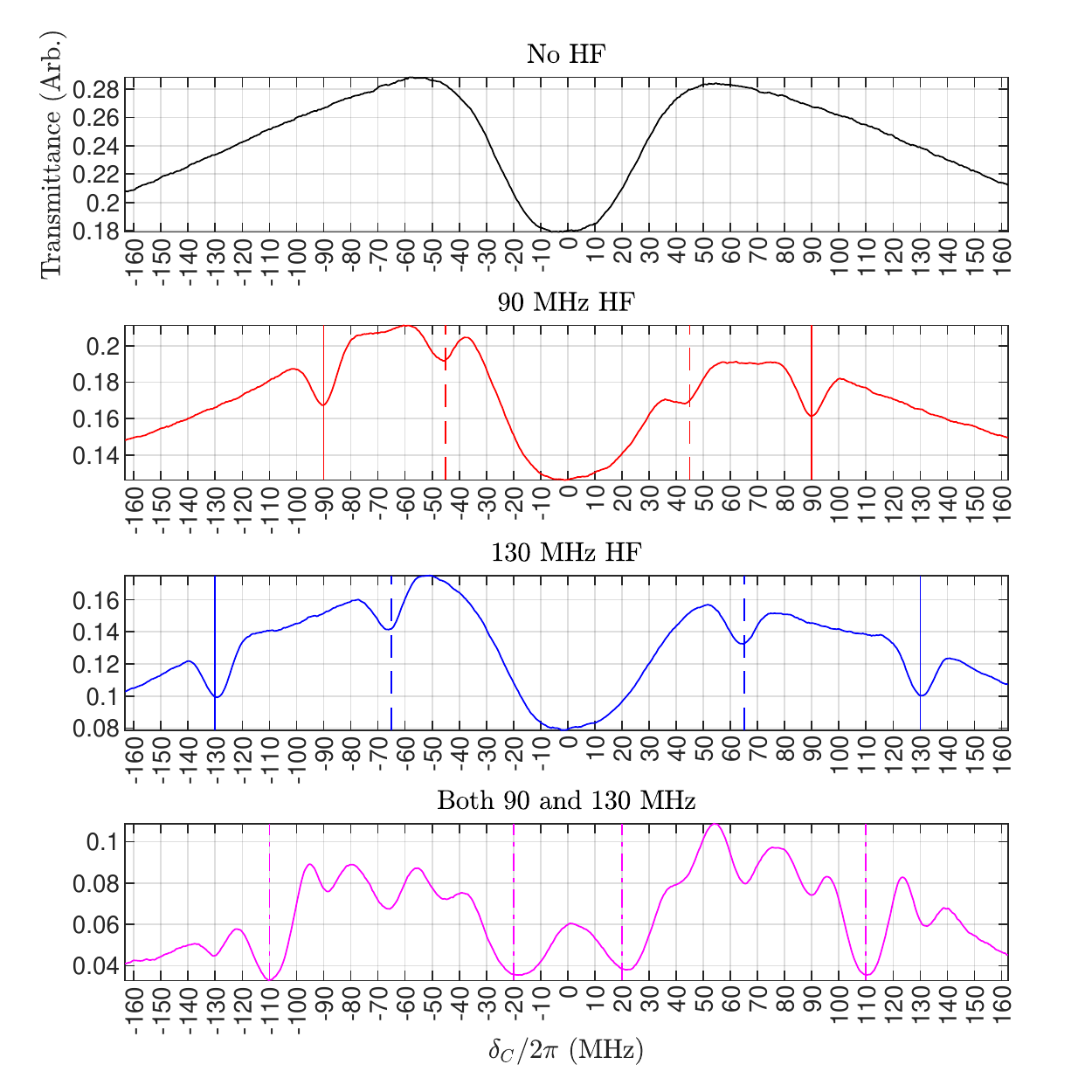}
    \put(14,89){(a)}
    \put(14,66){(b)}
    \put(14,42.2){(c)}
    \put(14,20){(d)} \end{overpic}
    \caption{Demonstration of multiple simultaneous tones.
    We average spectra over a wide range of $\Omega_{dr}$, and remove the Stark offset. 
    We compare (a) no HF signal, (b) a 90~MHz, (c) 130~MHz HF field, and (d) both fields, each at $P_{HF}=-10$~dBm($\approx 0.14$~V/cm at 100~MHz). 
    We denote the $\pm2$ and the $\pm1$ absorption features by solid lines and dashed lines, respectively, for the single tones (b,c). 
    When both signals are applied in (d), we denote additional features at half the sum (solid) and difference (dashed) frequencies.}
    \label{fig:multitone}
\end{figure}

\section{Applications}

We demonstrate two applications for HF field detection using the approach presented here. The first example demonstrates the ability to detect multiple low frequencies fields and the second example demonstrate the detection of a amplitude modulation (AM) of a HF carrier.

\subsection{Multiple Signal Detection}

First, we demonstrate ability to detect multiple HF field simultaneously, and the transmitted spectra for multiple simultaneous HF fields are shown Fig.~\ref{fig:multitone}. 
This figure shows the spectra averaged over many values of $\Omega_{dr}$ (as if collapsing the plots of Fig.~\ref{fig:ScanDressing} vertically).
We compare (a) no HF field, (b) a 90~MHz, and (c) a 130~MHz field separately, then (d) both simultaneously. 
For illustration, we subtract the Stark shift so the avoided crossings are symmetric around zero.
Notably, we observe Floquet states at the sum and difference frequencies, which then appear as absorption dips, noticeably  stronger than either single signal. 
% These additional resonances complicate this method's application as a general spectrum analyzer, but introduce interesting phase coherence and multi-photon schemes to be explored. 

\begin{figure*}
    \centering
    \begin{overpic}[width=0.57\textwidth]{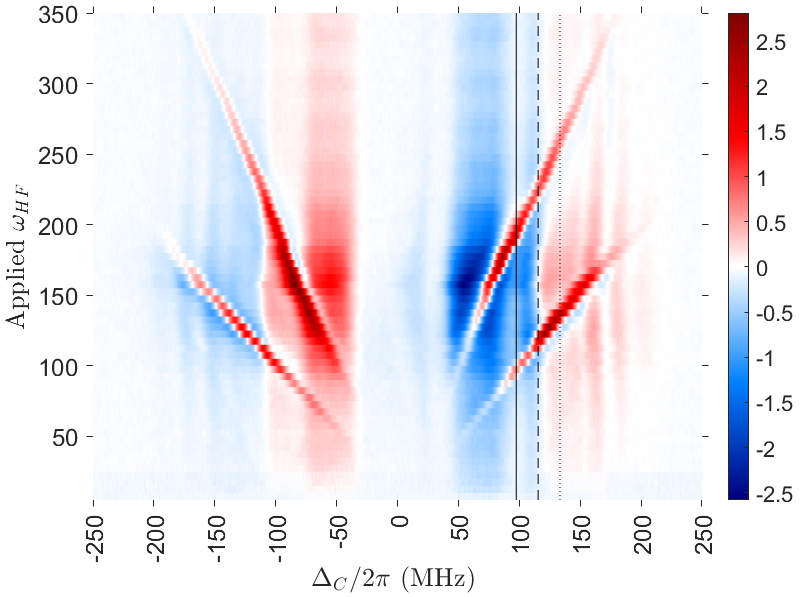}
    \put(-2,68){\large (a)}
    \put(16,33){\large $-2$}  \put(75,33){\large $+2$}
    \put(20,60){\large $-1$}  \put(75,60){\large $+1$}
        \end{overpic} 
    \begin{overpic}[width=0.4\textwidth]{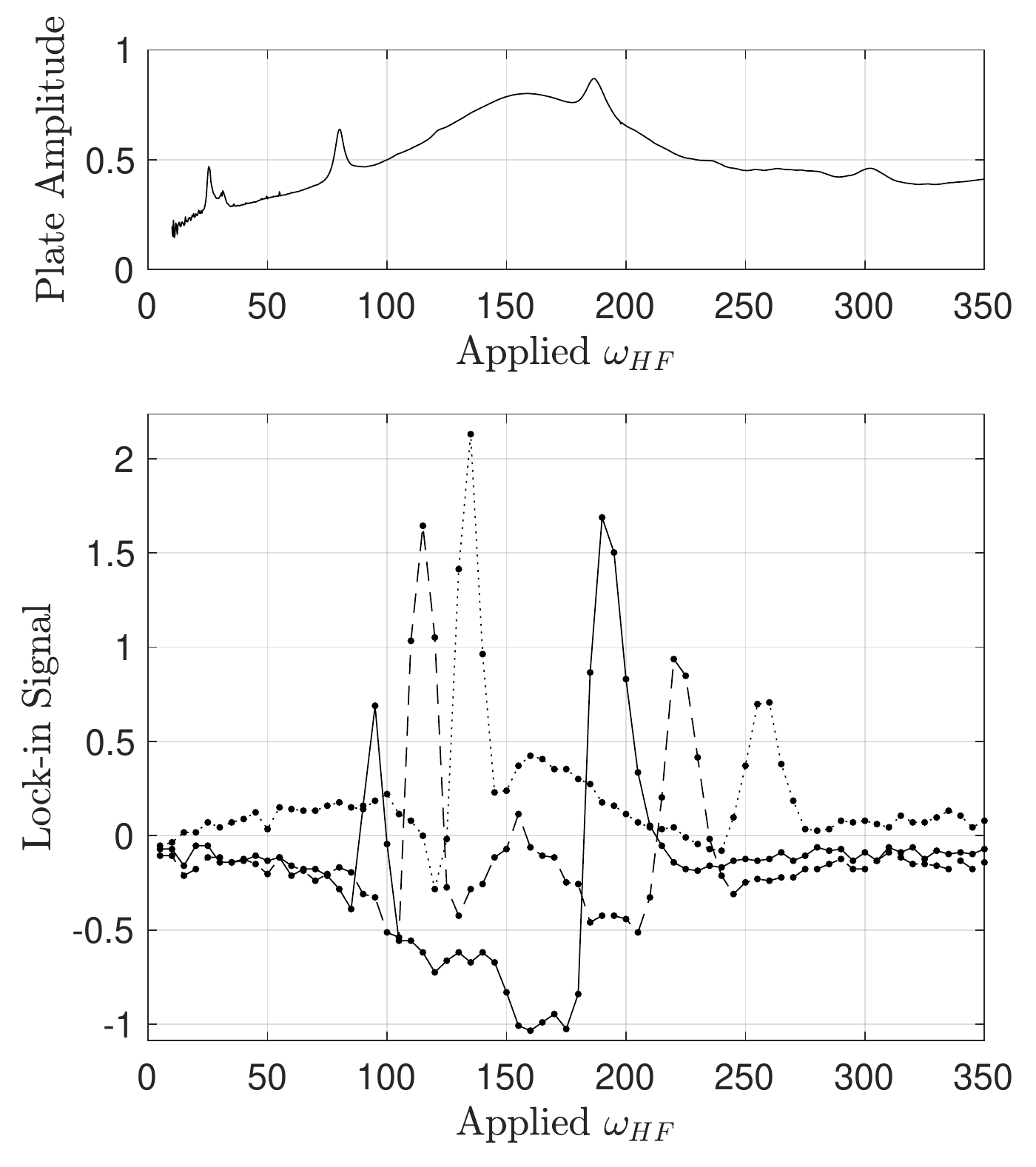}
      \put(15,89){\large (b)}
      \put(15,55){\large (c)} 
    \end{overpic}
    \caption{    \label{fig:signals}
    Detection of AM signals across HF carrier frequency. 
    (a) We plot in-phase lock-in amplifier of the laser transmittance signal referenced to a 10~kHz modulation of $P_{HF}$, in a waterfall over  $\omega_{HF}$, 
    for fixed $\Omega_{dr}\approx2\pi\cdot100$~MHz. 
    (b) Expected relative field amplitude from plate impedence measurements
    (c) Corresponding vertical slices of lock-in signal, as if fixing $\Delta_C$ and $\Omega_{dr}$}
\end{figure*}

\subsection{HF Carrier AM Detection}\label{sec:signals}
The ability to continuously measure amplitude of an HF carrier via a laser power transmittance allows for real-time, down-converted reception of amplitude modulated (AM) signals. 
Notably, this cm-sized atom sensor package is capable of simultaneous reception of meters-long wavelength carriers and microwave frequencies.
To demonstrate AM reception (modulating $E_{HF}$), we apply at 10 kHz AM signal to a varying the $\omega_{HF}$ carrier frequency while $\Omega_{dr}$ is providing a significant splitting.
We then use a lock-in amplifier on the laser transmittance signal, referenced to the AM signal, while scanning $\delta_C$, tuning lock-in phase to optimize positive peaks.
In-phase peaks in the spectrum indicate resonant reception is possible by tuning $\Omega_{dr}$ and $\delta_C$. 
In Fig.~\ref{fig:signals}(a), we show in-phase lock-in amplifier signal as the color axis, across scanning coupling laser frequency, in a waterfall over $\omega_{HF}$, with $\Omega_{dr}/2\pi$ held fixed near $200$~MHz.
As a result of the induced splitting central to this work, we observe in-phase modulation of the transmittance signal in both AT peaks, when both $\pm1\omega_{HF}$ and $\pm2\omega_{HF}$ are brought resonant with $\Omega_{dr}$. 
We observe signals using only $\approx$20~$\mu$W carrier power, with 50\% amplitude modulation, and notably, the Floquet sidebands alone are not visible outside the parts of the spectrum populated by AT peaks. 
For constant power, the electric field $E_{HF}$ from the plate's impedance depends on $\omega_{HF}$, and we plot $\sqrt{1-10^{S_{11}/10}}$, using reflection coefficient of the plates $S_{11}$ in dB, measured by a SNA in Fig.~\ref{fig:signals}(b).
We also observe the intensity-based Stark shifting \cite{liu2022highly} of all EIT peaks in the absence of a resonant $\omega_{HF}$ carrier, visible in the broad in- and out-of-phase coloring on the right and left sides of each AT peak, around $\delta_C\approx\pm100$.

In Fig.~\ref{fig:signals}(c), we plot lock-in data slices to simulate the tuned in or `pseudo-resonant' case by fixed laser detuning $\delta_C$ (and $\Omega_{dr}$), and scanning $\omega_{HF}$ with 5~MHz steps. 
Resonant peaks in the lock-in data demonstrate the ability to pick out particular carrier frequencies using laser detuning alone.
From Fig.~\ref{fig:signals}(c), we can estimate linewidth of order 10~MHz to 20~MHz, using a 5~MHz step size, and this will be the subject of future study. 
% although other experimental parameters makes it difficult to determine for a general case. 

\section{Conclusion}\label{sec:conc}
We have demonstrated a method to detect relatively low frequency (3~MHz to 300~MHz) fields using Rydberg states of atoms, by driving AT splitting with Rabi frequency approximately matching the HF frequency, and its second harmonic. 
The low end of the frequency range is set by EIT linewidth, and the upper end is set by dressing field uniformity.
The demonstrated dynamic range is on par with previous TM sideband measurements \cite{bason2010enhanced,miller2016radio,jiao2017atom}, although could be improved with a LO, or other methods. 
The observation of a second non-linear splitting due to $E_{HF}$ enables a path to measure field intensity in the HF and VHF range, as well as dc field components, and the frequency of multiple fields simultaneously, from atomic transmittance spectra. 
While the two-level Floquet theory comparison was able to capture many of the spectral features observed, we expect the precision of this style of field measurement can be improved using a multi-state Stark-shifted approach, especially at higher field values. 

Looking toward applications, we demonstrated signal detection on HF carriers, demonstrating its effective band-pass through Rabi matching, which is unavailable to previous Stark-based reception. 
One apparent limitation of the current apparatus is the non-uniform $\Omega_{dr}$, which will be improved upon in future experiments. 
% Interestingly, we found that a slight splitting of the EIT line caused an increase in sensitivity by converting amplitude modulation of differential Stark shifts into an effective atomic frequency modulation. 
As a receiver, this compact cm-sized atomic vapor cell is significantly smaller than wavelength-scale antennas, yet enables a quantum receiver for HF and VHF signals. 

\appendix
\section{Two-level Floquet Theory}\label{appx:theory}
When the oscillating HF field modulates state energy faster than the linewidth and away from atomic energy resonances with allowed transitions, one observes induced quasi-energy states at spacing of $\hbar\omega_{HF}$, forming an energy ladder $N\hbar\omega_{HF}$ for integer $N$.
Floquet theory prescribes a method to expand the basis states of a time-dependent Hamiltonian ($i\hbar\frac{\textrm{d}c}{\textrm{d}t}=H(t)c(t)$, where $c(t)=\{d(t),f(t)\}$, the time-evolving state coefficients), into an infinite ladder of states
$$c(t) = \sum_{N=-\infty}^{\infty} c_N~ \textrm{exp}\left(-i\varepsilon_Nt/\hbar \right)  $$
Each new basis state has quasi-energy $\varepsilon_{N} = \varepsilon_{0} + N\hbar\omega_{HF}$, relative to the time-average state energies at $\varepsilon_{D/F,0}/\hbar = \omega_{D/F} + \Sigma_{D/F}^- + \Sigma_{D/F}^\sim$.
Population probabilities are calculated by mapping the initial $|d_0|^2=1$ state into a new diagonalized basis, and normalizing the resulting eigenvectors. 
The coefficients of the the $D$ and $F$ states' $N^\textrm{th}$ sideband are given by $d_N$ and $f_N$ respectively, which are rendered time-independent by this approach. 
In the absence of a dc field, this `ladder' of energy states has spacing $2N\omega_{HF}$, as has been observed in Rydberg atoms \cite{bason2010enhanced, miller2016radio, jiao2016spectroscopy, miller2017optical, jiao2017atom,  paradis2019atomic}, well-predicted by this Floquet theory.
In the limiting case of $E_{dc}=0$ and $N\rightarrow\pm\infty$ Floquet states, the height of the $N^\textrm{th}$ sideband is given as Bessel functions of order $N$:
$J_N^2\left( \frac{-\alpha E_{HF}^2}{8\omega} \right)$. 
Calculating a finite Floquet basis requires a truncation of states (from $-N_{max}$ to $N_{max}$, we use $N_{max}=24$ throughout) for both the $D$ and $F$ ladders, and for our purposes must include the rf dressing and the dc field.

The time-dependent terms of Eq.~\ref{eq:terms} will time-average to $0$, so we have $\langle\mathcal{E}^2\rangle = E_{dc}^2 + \frac{E_{HF}^2}{2}$, using the RMS value for the ac part. 
With this field present, the $D$ and $F$ states shift on average by
$\Sigma_{D}^- + \Sigma_{D}^\sim>0$, 
and $\Sigma_{F}^- + \Sigma_{F}^\sim<0$.
The Stark shift in both states brings the applied dressing frequency $\omega_{dr}$ down to the shifted resonance with $\delta_{dr}$ (defined in Sec.~\ref{sec:The}) kept near zero as the fields increase.
For these states, this differential shifting drops the observed resonance by $(\alpha_D - \alpha_F)/2 = -7.552$~GHz per (V/cm)$^2$. %although our experimental value differed significantly, in indicating smaller values of $\alpha_F$ than the theoretical average-field value used here.

Applying rf fields nearly resonant to an allowed atomic transition (here, $\omega_{dr}\simeq\omega_F-\omega_D$), one can determine energy levels of the atom-photon-interaction using dressed atom theory. 
The two-level coupling between the $\ket{D}$ and $\ket{F}$ state has dipole strength $\bra{56D_{5/2},m_J=\frac{1}{2}}e\cdot\hat{z}\ket{54F_{5/2},m_J=\frac{1}{2}} = \wp_{DF}\approx1746~ea_0\approx2234$~MHz per (V/cm).
The time-dependent  two-level Hamiltonian is
\begin{equation}\label{eq:timeHam}
\begin{split}
&H(t)/\hbar=\\
&\left( {\begin{array} {cc} 
\omega_D -\alpha_D\mathcal{E}^2(\omega_{HF}t)/2 +\omega_{dr} & \Omega_{dr}\cos(\omega_{dr}t) \\
\Omega_{dr}\cos(\omega_{dr}t) &  \omega_F -\alpha_F\mathcal{E}^2(\omega_{HF}t)/2
\end{array} } \right) 
\end{split}
\end{equation}
using shorthand $\mathcal{E}^2(\omega_{HF}t) = \left( E_{dc} + E_\textrm{HF}\cos(\omega_{HF}t)\right)^2 $.
Taking the time-average Stark shifts, we have a well-defined $\delta_{dr}$, and using the rotating wave approximation (RWA) to shortcut another Floquet basis, we write the typical dressed atom Hamiltonian for the two-level atom-photon basis $\{\ket{D,(N_{dr}+1)\omega_{dr}}, \ket{F,N_{dr}\omega_{dr}}\}$ and setting $\omega_D=0$:
\begin{equation}\label{eq:timeHam}
H_{atom-photon}= \hbar \left( {\begin{array} {cc} 
0 & \Omega_{dr}/2 \\
\Omega_{dr}/2 &  -\delta_{dr}
\end{array} } \right) 
\end{equation}
Driving this transition resonantly ($\delta_{dr}=0$), we induce AT splitting linear with the field applied, given by the Rabi frequency: $\Omega_{dr}=\wp_{DF}E_{dr}/\hbar$.
Off resonance, eigen-energies are given by 
$\varepsilon_{0,\pm} = \frac{\hbar}{2}\left( -\delta_{dr} \pm \sqrt{\delta^2_{dr} + \Omega_{dr}^2} \right) $ 
noting that $\delta_{dr}$ is defined including time-averaged shifts, and therefore $\omega_{dr}$ must shift with $P_{HF}$ to maintain $\delta_{dr}\simeq0$.
The dressing field forms an effective two-level system for each of pair of quasi-energy states, connecting each $d_N$ component to the same $N$'s $f_N$ state, splitting either state's $\varepsilon_N$ quasi-energy into $\varepsilon_{N,\pm}$ dressed state components.

Looking to the time-dependent terms of Eq.~\ref{eq:terms}, we have rotating components at $\omega_{HF}$ with strength $\Sigma_{D/F}^\times/2$, the dc mixing term, and at $2\omega_{HF}$ with strength $\Sigma_{D/F}^\sim/2$, the pure ac term.
These terms connect Floquet states with $\Delta N=\pm1$, and $\Delta N=\pm2$, in addition to the dressing between the $d_N$ and $f_N$ states.
We can write the the Schr\"odinger equation as a system of linear equations:
\begin{equation}
\begin{split} 
\varepsilon d_N =& (\Sigma_D^-+\Sigma_D^\sim +\omega_{dr} +N\omega_{HF})~d_N \\
&+ \Sigma_D^\times/2~(d_{N+1} +  d_{N-1} ) \\
&+ \Sigma_D^\sim/2~(d_{N+2} +  d_{N-2} )\\
&+\Omega_{dr}/2~f_N 
\end{split}
\end{equation}

and

\begin{equation}
\begin{split} 
\varepsilon f_N =& (\Sigma_F^-+\Sigma_F^\sim +N\omega_{HF})~f_N \\
&+ \Sigma_F^\times/2~(f_{N+1} +  f_{N-1} ) \\
&+ \Sigma_F^\sim/2~(f_{N+2} +  f_{N-2} )\\
&+\Omega_{dr}/2~d_N 
\end{split}
\end{equation}

The resulting Hamiltonian and basis state vector for this dressed-Floquet system is given in Eq.~\ref{eq:bigHam}, using $\omega_D+\Sigma_D^-+\Sigma_D^\sim $ as the 0 energy reference.
% in the basis 
% \begin{equation} 
% \begin{split} 
% &\{ d_{-N_{max}}, ... ,d_{-2},d_{-1},d_{0},d_{1},d_{2}, ...,d_{+N_{max}},\\
% &f_{-N_{max}}, ... ,f_{-2},f_{-1},f_{0},f_{1},f_{2}, ...,f_{+N_{max}}   \}^T
% \end{split}
% \end{equation}
The diagonal blocks include on the diagonal the quasi-energy ladder, shifted by  time-averaged fields and rendered near-degenerate between diagonal blocks by the addition of $\hbar\omega_{dr}$, included in $\delta_{dr}\simeq0$.
Just off the diagonal, we have the $\omega_{HF}$ and $2\omega_{HF}$ interactions between (near-) neighboring quasi-energies. 
In the off-diagonal blocks, we have the dressing coupling $\Omega_{dr}$ between corresponding near-resonant states.
\begin{widetext}
\begin{equation}\label{eq:bigHam} \footnotesize
\begin{split}
&\frac{\hat{\mathcal{H}}}{\hbar} \left(\begin{array}{c}
d_N \dots\\ \hline f_N\dots
\end{array}\right)
=  \varepsilon/\hbar
\left(\begin{array}{c}
d_N \dots\\ \hline f_N\dots
\end{array}\right) = \\
&\left( \begin{array} {ccccccc|ccccccc} 
\ddots&\Sigma^\times_D/2&\Sigma^\sim_D/2&0&0&0&\ddots& \ddots&0&0&0&0&0&\ddots \\
\Sigma^\times_D/2&-2\omega_{HF}& \Sigma^\times_D/2&\Sigma^\sim_D&0&0&0&0&\Omega_{dr}/2&0&0&0&0&0\\
\Sigma^\sim_D/2 &\Sigma^\times_D/2&-\omega_{HF}& \Sigma^\times_D/2&\Sigma^\sim_D/2&0&0&0&0&\Omega_{dr}/2&0&0&0&0\\
0&\Sigma^\sim_D/2&\Sigma^\times_D/2& 0& \Sigma^\times_D/2&\Sigma^\sim_D/2&0&0&0&0&\Omega_{dr}/2&0&0&0\\
0&0&\Sigma^\sim_D/2&\Sigma^\times_D/2&\omega_{HF}& \Sigma^\times_D/2&\Sigma^\sim_D/2 &0&0&0&0&\Omega_{dr}/2&0&0\\
0&0&0&\Sigma^\sim_D/2&\Sigma^\times_D/2&2\omega_{HF}& \Sigma^\times_D/2&0&0&0&0&0&\Omega_{dr}/2&0\\
\ddots&0&0&0&\Sigma^\sim_D/2&\Sigma^\times_D/2&\ddots&\ddots&0&0&0&0&0&\ddots\\
\hline
\ddots&0&0&0&0&0&0&\ddots&\Sigma^\times_F/2&\Sigma^\sim_F/2&0&0&0&\ddots\\
0&\Omega_{dr}/2&0&0&0&0&0&\Sigma^\times_F/2& -2\omega_{HF}-\delta_{dr}& \Sigma^\times_F/2&\Sigma^\sim_F&0&0&0\\
0&0&\Omega_{dr}/2&0&0&0&0&\Sigma^\sim_F/2 &\Sigma^\times_F/2& -\omega_{HF}-\delta_{dr}& \Sigma^\times_F/2&\Sigma^\sim_F/2&0&0\\
0&0&0&\Omega_{dr}/2&0&0&0&0&\Sigma^\sim_F/2&\Sigma^\times_F/2& -\delta_{dr}& \Sigma^\times_F/2&\Sigma^\sim_F/2&0\\
0&0&0&0&\Omega_{dr}/2&0&0&0&0&\Sigma^\sim_F/2&\Sigma^\times_F/2& \omega_{HF}-\delta_{dr}& \Sigma^\times_F/2&\Sigma^\sim_F/2 \\
0&0&0&0&0&\Omega_{dr}/2&0&0&0&0&\Sigma^\sim_F/2&\Sigma^\times_F/2& 2\omega_{HF}-\delta_{dr}& \Sigma^\times_F/2\\
0&0&0&0&0&0&\ddots&\ddots&0&0&0&\Sigma^\sim_F/2&\Sigma^\times_F/2&\ddots\\
\end{array}  \right) 
\left(\begin{array}{c}
\vdots\\d_{-2}\\d_{-1}\\d_{0}\\d_1\\d_2\\\vdots\\\hline\vdots\\f_{-2}\\f_{-1}\\f_0\\f_1\\f_2\\\vdots
\end{array}\right)
\end{split}
\end{equation}
\end{widetext}

\section{Determining Effective Polarizability}\label{sec:pol}
\begin{figure}
    \centering
    \includegraphics[width=\columnwidth]{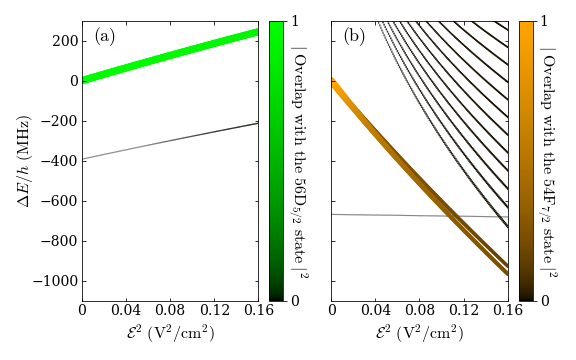}
    \caption{Results from a many-level calculation of Stark energy shift. Fitting across electric field values gives the effective polarizability used in this paper for $\ket{D}$ in (a), and $\ket{F}$ in (b).}
    \label{fig:polcalc}
\end{figure}

A true Stark energy shift calculation \cite{zimmerman1979stark} considers coupling to all other atomic states, but in the low field limit, an effective quadratic polarization coefficient is fit from this process.
Rather than use the low-field limit for polarizability, we calculate effective polarizabilities in-house, given slightly non-quadratic behavior, for fields up to $\approx0.4$~V/cm.
In Fig.~\ref{fig:polcalc}, we demonstrate an example of calculating the Stark shift from a dc electric field when including all the states which significantly contribute to the energy shift. 
In principle, this effective polarizability value changes for different maximum fields, which alters the implied measurements of electric field from a spectral gap. 
% Due to the difference between the low-field polarizability and at the fields used in this work, we give a nominal 10\% error budget on intensity values derived from these constant values of polarizability.

\section{List of Symbols}\label{appx:abbrev}

\begin{table*}[h!]
\setlength{\tabcolsep}{20pt} % Default value: 6pt
\renewcommand{\arraystretch}{1.5} % Default value: 1
\centering
\begin{tabular}{|c|c|} 
 \hline 
 Symbol & Meaning  \\ 
\hline\hline
 EIT  & Electromagnetically-Induced Transparency \\ \hline 
  AT  & Autler-Townes effect, dressed state splitting \\ \hline 
  HF  & High Frequency radio band, 3~MHz to 30~MHz \\ \hline 
  VHF  & Very High Frequency radio band, 30~MHz to 300~MHz \\ \hline 
 TM & Townes-Merritt effect, Floquet quasi-energy sidebands \\ \hline 
 $\delta_{C}$ & Coupling laser detuning from low-field (reference cell) EIT resonance \\ \hline 
 $\omega_{dr}$ & Dressing field angular frequency \\ \hline 
 $\Omega_{dr}$ & HF field angular frequency \\ \hline 
 $P_{HF}$ & Power applied to low frequency plates\\ \hline
 $E_{HF}$ & Low frequency electric field on plates\\ \hline
 $E_{dc}$ & Static electric field inside vapor cell\\ \hline
 $\mathcal{E}(t)$ & Time-evolving total electric field seen by the atoms\\ \hline
 $(~\cdot~)_{D/F}$ & Two values represented, for 56D$_{5/2}$ and 54F$_{7/2}$\\ \hline
 $\hbar\omega_{D/F}$ & Energy of the unperturbed \{D,F\} state \\ \hline 
 $\alpha_{D/F}$ & Polarizability of the \{D,F\} state \\ \hline
 $\hbar\Sigma_{D/F}^-$ & Energy shift from dc + RMS electric field of the \{D,F\} state. See Eq.~\ref{eq:sigmadef} \\ \hline
 $\hbar\Sigma_{D/F}^\sim$ & Energy shift from the HF RMS electric field of the \{D,F\} state. See Eq.~\ref{eq:sigmadef} \\ \hline
 $\hbar\Sigma_{D/F}^\times$ & Energy shift from ac/dc cross term of the \{D,F\} state. See Eq.~\ref{eq:sigmadef} \\  \hline
 $\delta_{dr}$ & Detuning of $\omega_{dr}$ from the shifted $\ket{D}$ and $\ket{F}$ state energy gap \\ \hline 
 $S_{\delta}$ & Scaling parameter for $\delta_{dr}$, near 1.0 \\ \hline
 $\varepsilon_N$ & N$^\textrm{th}$ Floquet sideband quasi-energy \\ \hline
 $(~\cdot~)_{\pm}$ & Two values represented, for the upper ($+$) and lower ($-$) AT dressed state\\ \hline
\end{tabular}
\caption{Table of non-standard variables used in this paper. }
\label{table}
\end{table*}

\bibliography{lowfreq}

\end{document}